# Nearly Exclusive Growth of Small Diameter Semiconducting Single-Wall Carbon Nanotubes from Organic Chemistry Synthetic End-Cap Molecules

Bilu Liu[1], Jia Liu[1], Hai-Bei Li[2], Radha Bhola[3], Edward A. Jackson[3], Lawrence T. Scott[3], Alister Page[4], Stephan Irle[5], Keiji Morokuma[6], Chongwu Zhou[1]*

1. Department of Electrical Engineering and Department of Chemistry, University of Southern California, Los Angeles, California, 90089, United States

2. School of Ocean, Shandong University, Weihai 264209, China

3. Merkert Chemistry Center, Boston College, Chestnut Hill, Massachusetts 02467, United States

4. Newcastle Institute for Energy and Resources, The University of Newcastle, Callaghan 2308, Australia

5. WPI-Institute of Transformative Bio-Molecules (ITbM) & Department of Chemistry, Graduate School of Science, Nagoya University, Nagoya 464-8602, Japan

6. Fukui Institute for Fundamental Chemistry, Kyoto University, Kyoto, 606-8103, Japan

Corresponding author chongwuz@usc.edu (C. Z.)

**Abstract**

**The inability to synthesize single-wall carbon nanotubes (SWCNTs) possessing uniform electronic properties and chirality represents the major impediment to their widespread applications. Recently, there is growing interest to explore and**

synthesize well-defined carbon nanostructures, including fullerenes, short nanotubes, and sidewalls of nanotubes, aiming for controlled synthesis of SWCNTs. One noticeable advantage of such processes is that no metal catalysts are used, and the produced nanotubes will be free of metal contamination. Many of these methods, however, suffer shortcomings of either low yield or poor controllability of nanotube uniformity. Here, we report a brand new approach to achieve high efficiency metal-free growth of nearly pure SWCNT semiconductors, as supported by extensive spectroscopic characterization, electrical transport measurements, and density functional theory calculations. Our strategy combines bottom-up organic chemistry synthesis with vapour phase epitaxy elongation. We identify a strong correlation between the electronic properties of SWCNTs and their diameters in nanotube growth. This study not only provides material platforms for electronic applications of semiconducting SWCNTs, but also contributes to fundamental understanding of the growth mechanism and controlled synthesis of SWCNTs.

**Introduction.**

Single-wall carbon nanotubes (SWCNTs) represent attractive materials for the next generation nanoelectronics, macroelectonics, and optoelectronics, owing to their intrinsic small dimensions, excellent electronic and optical properties, chemical inertness, mechanical robustness, and other outstanding properties.[1, 2] To transform the electronics applications of SWCNTs from a sought-after dream goal to a high-impact reality, the electronic properties of SWCNTs must be precisely controlled.[3-5] It has been well documented that SWCNTs can be either metals or semiconductors, depending critically on their geometrical structures, or specifically, their chirality.[2] Currently, researchers believe that the chirality, and therefore the electronic properties, of a SWCNT become fixed during the initial nucleation step and that the follow-up steady growth stage will not change that chirality, but will just extend the nanotube length. This is supported by the fact that ultra-long SWCNTs typically possess the same chirality along their entire length, unless the growth conditions change.[6] Therefore, nanotube nucleation control at the initial stage is the key to solving the structure and property heterogeneity problem.

Metal nanoparticles with sizes of only a few nanometers are traditional catalysts for the synthesis of SWCNTs, and nanotubes with narrow heterogeneity have been successfully grown with varying degree of success in the past decade.[7-18] On the other hand, there has been increasing interest in recent years in using structurally well-defined carbon nanomaterials[19-28] to initiate nanotube growth, with an aim of producing uniform SWCNTs. Equally importantly, the potential influence of metal contamination on the properties and applications of nanotubes will be eliminated in such metal-free growth systems. For example, nanotube cloning has been demonstrated by Zhang, Liu and co-workers[24] and by our own group[25, 26] recently, with the newly grown SWCNT segments having the same chirality as the seeds employed, based on Raman spectroscopic analysis. Nevertheless, the yield of the cloning process is still rather low at the current stage, largely due to very low areal number densities of nanotube seeds. On the other hand, fullerenes[29, 30] and sidewall rings of nanotubes[31] have been reported to grow SWCNTs very recently, but the products are mixtures of metallic and semiconducting SWCNTs. Here, we report a new nanotube growth method that combines bottom-up organic chemistry synthesis with an elongation method resembling vapour phase epitaxy (VPE) to achieve metal-catalyst-free growth of nearly pure semiconducting SWCNTs and their aligned arrays. We used pure, structurally well-defined molecular end-caps of nanotubes, which are completely metal-free, to initiate further nanotube growth; scanning electron microscopy (SEM) and atomic force microscopy (AFM) characterization confirm the growth of horizontally aligned SWCNTs with high yield, indicating the high efficiency of the molecular end-caps. Furthermore, Raman spectroscopic analysis with multiple lasers indicates the very small diameters of the as-grown SWCNTs, and single nanotube and nanotube array field-effect transistor (FET) measurements unambiguously confirm the nearly exclusive growth of semiconducting SWCNTs, with purity higher than 97%, one of the highest purity so far from a direct growth strategy. The mechanism of SWCNT growth from carbon nanostructures, including the relationship between seed size and nanotube diameter, chirality-evolution process of SWCNTs, and the origin of the selective growth of semiconducting nanotubes are studied via experiments and density functional theory (DFT) calculations.

**Results**

We start with the corannulene molecule ($C_{20}H_{10}$, structure shown in Figure 1a) for the synthesis of the end-caps of nanotubes ($C_{50}H_{10}$, structure shown in Figure 1b), aiming at chirality-controlled growth of SWCNTs via molecular cap engineering. The synthesis details of end-caps can be found in the Method section. We point out metal elements such as Fe, Co, Ni, Cu, etc., which can act as catalysts for CVD growth of nanotubes, were not used in the above organic chemistry synthesis process. Therefore, it is safe to conclude that the as-formed $C_{50}H_{10}$ molecules are free of such metals, as confirmed by the energy dispersive X-ray (EDX) characterization (Figure S1). Later, we propose to use this nanotube-end-cap aiming for the chirality-controlled synthesis of SWCNTs via a strategy resembling VPE. The feasibility of VPE growth of SWCNTs has recently been demonstrated in our work using DNA-separated nanotube seeds.[25, 26]

The as-synthesized red-orange powder of $C_{50}H_{10}$ was first dissolved in toluene to make a stable solution (Figure S2). Then, the quartz substrates with $C_{50}H_{10}$ molecules were subjected to a horizontal CVD furnace for subsequent nanotube growth (See Methods for CVD details). We initially conducted nanotube growth experiments without any pretreatment of $C_{50}H_{10}$. However, no nanotubes grew under broad growth conditions (Figure S3). Instead, we observed two typical features on the quartz surface after the CVD process, i.e., a clean surface without anything grown on it or a dirty surface with dense amorphous carbon deposits (Figure S3). Such amorphous carbon deposition was typically observed for growth conditions with either high $CH_4$ and $C_2H_4$ partial pressures or high growth temperatures, which result in considerable thermal pyrolysis of carbon sources, as evidenced by the blackening of the quartz reaction tube. Next, we reexamined the whole process and considered the possibility that rims of the $C_{50}H_{10}$ molecules might be covered by other $C_{50}H_{10}$ molecules or solvents, embedding the active edges inside larger aggregates, as supported by the AFM characterization (shown later). Faced with this conjecture, we speculated that pretreatment might be necessary to initiate nanotube growth from this molecular end-cap. After extensive exploration, ultimately, we found that high temperature air treatment, followed by water vapour treatment, is very effective in activating $C_{50}H_{10}$ for nanotube growth. In particular, we learned that air oxidation at 500 °C followed by water vapour treatment at 900 °C gives the highest nanotube yield (see Methods, Figure 2 and Figure S4).

We used SEM to examine the overall growth efficiency of nanotubes from pretreated $C_{50}H_{10}$. The phenomenon of the coffee ring effect clearly demonstrates that nanotubes were indeed grown from pretreated $C_{50}H_{10}$ molecules. To demonstrate this, we deposited ~5 μl of $C_{50}H_{10}$ solution in toluene onto a quartz substrate and allowed it to nearly dry under ambient conditions. During the drying process, most of the $C_{50}H_{10}$ molecules were left on the boundary of the solvent, due to capillary force.[32] The inset of Figure 2a shows a digital camera image of such a substrate, in which the red-orange deposit of $C_{50}H_{10}$ molecules localized mostly along a circle can be clearly discerned. Figure 2a shows a low magnification SEM image of this substrate after the nanotube growth process, using $C_{50}H_{10}$ pretreated in air at 500 °C and water vapour at 900 °C prior to nanotube growth; a bright circle-shaped area is visible. Zoom-in SEM images of the square areas b and c (Figure 2b and Figure 2c) clearly show the highly efficient growth of dense SWCNTs. The yield of SWCNTs grown from $C_{50}H_{10}$ is much higher than that grown from short nanotube seeds, since the area number density of small $C_{50}H_{10}$ molecules is significantly larger than the area number density of nanotube seeds. High magnification SEM characterization shows that the nanotubes are aligned along the quartz surface (Figure 2d). We also found nanotubes at relatively low density inside the coffee ring boundary (area e in Figure 2a), derived from the small amount of $C_{50}H_{10}$ molecules deposited inside the coffee ring (Figure 2e). Some nanotubes grown at high density areas are less aligned and bent (Figures 2b-2d), because the existing molecular clusters or other nanotubes on substrate can result in changes of the growth directions of nanotubes. In contrast, the nanotubes grown at low density areas are typically straight and align along the crystalline orientation of quartz (Figure 2e). As control experiments, we performed nanotube synthesis experiments using blank quartz substrates, following the identical air and water pretreatment, and no SWCNTs growth was observed (Figure S5). Overall, the above experiments unambiguously demonstrate that nanotube growth is indeed initiated by the deposited $C_{50}H_{10}$ molecules. We used AFM to study the diameters of the as-grown SWCNTs, and found that most nanotubes have heights below 1 nm. For example, Figure 2f shows a representative AFM image of a SWCNT with a diameter of ~0.6 nm. We have found, however, that the as-grown SWCNTs contain some bundles, which introduces uncertainty with the use of AFM for nanotube diameter measurements.

To analyse further the diameter, chirality, and quality of the SWCNTs grown from the pretreated $C_{50}H_{10}$ molecular end-caps, we performed systematic Raman spectroscopic analysis with multiple lasers. We found that the $C_{50}H_{10}$ molecules themselves show very weak Raman signals under short laser wavelength (Figures S6). Figures 3a, 3b, and 3c show representative Raman spectra of as-grown SWCNTs with laser excitation wavelengths of 633 nm (Figure 3a), 514 nm (Figure 3b), and 457 nm (Figure 3c), respectively. The diameters of SWCNTs were deduced from the relationship between the frequencies of radial breathing modes (RBMs) in Raman spectra and nanotube diameters, using the equation $\mathbf{d}_t = \frac{223.5}{\omega_{RBM} - 12.5}$ . Here, $d_t$ is the diameter of nanotube in nanometer, and $\omega_{RBM}$ is the frequency of the RBM in cm$^{-1}$. We found this equation fits the diameter-RBM relationship in as-grown SWCNTs very well based on our previous studies.[25] The peaks marked with arrows are RBMs, while all the other peaks (indicated by asterisks) are from the quartz substrates (Figure S7). In these Raman spectra, most of the RBM peaks are located at wavelengths above 240 cm$^{-1}$, indicating that small diameter nanotubes, with $d_t$<1 nm, have been grown. Statistical analyses of the RBM frequencies of SWCNTs based on the three laser excitations are shown in Figures 3d, 3e, and 3f, and the diameter distribution of SWCNTs derived from the Raman characterizations are plotted in Figure 3g, which exhibits an average nanotube diameter of 0.82 nm.

We analysed the chirality information on the SWCNTs that was obtained by excitation with the three lasers used. Here, we emphasise that for such small diameter SWCNTs (<1 nm), Raman spectra are unambiguous with respect to chirality assignments, since adjacent SWCNTs have very distinct $E_{ii}$ values and RBM frequencies. Surprisingly, we discovered that most of the nanotubes are actually semiconductors, e.g., (8, 3), (6, 1), and (5, 1) in Figure 3a, (7, 3), and (5, 4) in Figure 3b, and (8, 1) in Figure 3c. More interestingly, we observed some RBMs at very high frequencies, for instance, the RBMs at ~517 cm$^{-1}$ are the peaks most frequently observed when using the 633 nm laser (Figures 3a and 3d), which corresponds to nanotube diameters of ~0.44 nm. We note that only (5, 1) SWCNT should show a RBM peak at this frequency. Therefore, we assign these RBMs to (5, 1) nanotubes. To the best of our knowledge, this is one of the smallest diameter SWCNTs ever reported to have been grown without the necessity of a template confinement.[33] We point out that the current growth method not only provides a practical way to synthesize such ultra-small diameter SWCNTs and produces valuable material platforms to allow studies of their exotic properties, but also provides critical information on the build-up of the electronic transition energy database of these small nanotubes, which is of fundamental importance to the study of structure-property relationships of SWCNTs and curvature-induced electronic property changes.[34] In addition, the very low defect-induced D-band to tangential G-band intensity ratio (<0.01) suggests a high quality of SWCNTs grown from the pretreated $C_{50}H_{10}$ molecules (Figure 3h and Figures S8).

Note that the above three lasers are not in resonance with the (5, 5) SWCNTs. We also used a 405 nm laser, which could excite (5, 5) SWCNTs, to characterize our sample. However, we did not observe any RBMs at ~340 $cm^{-1}$ related to (5, 5) nanotubes, indicating no or very small population of (5, 5) chirality in the sample. In fact, when using the 405 nm laser, we observed many fewer RBMs than when using the other three lasers (Figure S9 and Table S1). This is understandable since the 405 nm laser has high energy photons, and only very few nanotubes are in resonance with this laser based on the Kataura plot.

Raman spectroscopic characterization points to a trend that nanotubes grown from pretreated $C_{50}H_{10}$ may be enriched with semiconducting SWCNTs (Figure 3). It is difficult, however, to determine the precise metallic/semiconducting ratio of SWCNTs by Raman spectroscopy, owing to their resonant nature. To obtain a quantitative value for the proportion of semiconducting nanotubes rigorously, we performed systematic electrical transport measurements based on individual SWCNT FETs, as well as on aligned array SWCNT FETs combined with the electrical breakdown technique. We first transferred as-grown SWCNTs from quartz to Si/SiO$_2$ (90 nm oxide), using a polymer-mediated transfer process,[35] and fabricated bottom gate FET devices (Figure 4a, See Methods for the device fabrication and measurement details). We have fabricated a total of 13 chips and have tested more than 1000 devices, among which 147 working devices with nanotubes in the channel areas were identified. Among these devices, about 1/4 of them show individual SWCNTs in the channel, while the other 3/4 show parallel SWCNTs forming an array in the channel. Occasionally, a few devices were observed in which no SWCNTs were directly connected to the source and drain electrodes; instead, they formed a random network inside the channel. Such network devices are not included in the following discussion.

We plot the distribution of on/off ratios of individual SWCNT FETs (Figure 4b and Figure S10) and set an on/off ratio criterion of 10 to distinguish metallic SWCNTs from semiconducting ones.[10, 11, 36] The total number of individual SWCNT FETs is 34, and among them, 32 have on/off ratios larger than 10, giving a semiconducting SWCNT ratio of 32/34=94.1%. Figure 4c shows the typical transfer characteristics (I$_{DS}$-V$_G$) of a semiconducting SWCNT FET (SEM image of the device shown in the inset), which shows p-type behavior with an on/off current ratio of ~3.7 x $10^4$. As a comparison, Figure 4d shows the transfer characteristics of another individual SWCNT FET (SEM image shown in the inset) with an on/off ratio of ~7 (based on the red curve in Figure 3d with $V_D$=0.4 V). The output curves ($I_{DS}$-$V_D$) of the nanotube FETs shows linear behavior at small $V_D$ regimes (Figure S11), indicating the ohmic contact between the SWCNTs and the electrodes. Taking into account the evidence from Raman spectroscopic analysis with multiple lasers in Figure 3 that most of the nanotubes grown from pretreated $C_{50}H_{10}$ possess rather small diameters, we attribute the curves in Figures 4d to so called semi-metallic SWCNTs, such as (6, 3) SWCNTs, as detected by Raman characterization in Figure 3c. Such small-diameter semi-metallic SWCNTs have small but finite band gaps between their conduction band and their valance band in the electronic density of states and thus show gate dependence behavior,[37] as evidenced in Figures 4d. However, the band gaps of semi-metallic SWCNTs are typically very small, e.g., ~ 10 meV, which is reflected by the obvious ambipolar transport behavior observed in Figures 4d. Here we point out that the semi-metallic nanotubes with on/off ratio less than 10 (Figure 4d) are counted as metallic SWCNTs. Actually, we did not observe even a single device with on/off ratio of 1, which would correspond to a true metallic armchair SWCNT.

From SEM observations, we found that a large number of devices contain more than one SWCNT in the channel. For example, the SEM image in the inset of Figure 4e (also see Figure S12a) shows a four-SWCNT-array connected to the two electrodes. The transfer characteristic of this device, shown in Figure 4e, clearly demonstrates the semiconducting behavior, with an on/off ratio of ~520 for this nanotube-array-FET. The transfer characteristics of all such nanotube-array-FETs are summarized in Figure S13. We used the electrical breakdown technique[38] to count the actual number of SWCNTs for the device in Figure 4e. As shown in Figure 4f, we observed a total of three sudden decreases in $I_{DS}$ and, therefore, three SWCNTs being broken during the process. $V_D$ cannot be increased further since the gate oxide will be damaged at $V_D$ of ~80 V (Figure S14). We noted that after the breakdown of the 3[rd] SWCNT at $V_D$ of ~70 V, there is still current flow in the channel area, suggesting that at least one more SWCNT is still connected to the electrodes. Taking the SEM image (inset of Figure 4e and Figure S12a) and the breakdown experiments (Figure 4f) together, we conclude that there were originally a total of four semiconducting SWCNTs in this device, i.e., all the visible SWCNTs in the SEM image are indeed connected to both electrodes. This is reasonable, since we first transferred nanotubes and then conducted the electrodes deposition, putting the electrodes on top of the SWCNTs with good contact.

Similarly, we have performed systematically electrical breakdown experiments on nanotube-array FETs with on/off ratios <10. Figure 4g shows the transfer curves of such a device before and after electrical breakdown. The blue curve in Figure 4g shows the initial measurement of the device, which exhibits an on/off ratio of ~5. SEM inspection reveals two SWCNTs connected to both electrodes (Inset of Figure 4g and Figure S12b). After the electrical breakdown of the first metallic SWCNT at $V_G$= +5V (Figure 4h), the on/off ratio of the device increased to >4 x $10^3$, indicating only one metallic SWCNT in this device. Therefore, we conclude that this device contains one metallic SWCNT and one semiconducting SWCNT. Combining this electrical breakdown and counting technique with SEM imaging, as well as the individual SWCNT FET results, we have identified a total of 264 semiconducting SWCNTs and 8 metallic ones (Table S2), giving a semiconducting SWCNT ratio of 264/272=97.1%. The error for these statistics is given by equation $\delta = 1.96 \times \sqrt{\frac{\sigma^2}{N}} = 1.96 \times \sqrt{\frac{p(1-p)}{N}} = 2\%$. Here, $\delta$ is the statistic error, $\sigma$ is the standard deviation, N is the number of SWCNTs, p is the semiconducting SWCNT purity, and the confidence coefficient is set as 0.95. We note that this study is not only the very first example of the selective growth of semiconducting SWCNTs by a metal-free process, but also stands among one of the highest purity of semiconducting SWCNTs reported so far from a direct growth approach. Such small diameter semiconducting SWCNTs are preferred for short channel transistors since small diameter nanotubes exhibit much smaller OFF state current than large diameter ones.[39]

**Discussion**

So far, little is known about the growth mechanism of nanotubes from metal-free carbonaceous molecular seeds such as fullerenes,[29, 40] short nanotubes,[25, 26, 29] and carbon nanorings.[31] It is therefore important to investigate the mechanism of nanotube growth from the pretreated $C_{50}H_{10}$ molecular end caps, which will benefit further development of molecular seeds for structure-controlled nanotube growth. In this study, we focus on the following two major aspects: i) The relationship between the size of nanotubes and the sizes of molecular seeds from which nanotubes are grown from, ii) the underlying mechanism for the chirality-changed growth of SWCNTs and selective growth of semiconducting-predominated SWCNTs.

To shed some light on these issues, we first used AFM to study in detail the size evolution of deposited $C_{50}H_{10}$ clusters. AFM examination of the as-deposited $C_{50}H_{10}$ molecules shows an average particle size of 7.1 nm (Figure 5a and Figure S15a). Because the diameter of a single $C_{50}H_{10}$ molecule is approximately 1 nm,[21] this suggests aggregation of tens of $C_{50}H_{10}$ molecules into large clusters. It is obvious that such large molecular aggregates are not suitable for SWCNT growth, as observed in our experiments without seed pretreatment (Figure S3). After air and water vapour treatment at 500 °C and 900 °C, respectively, the sizes of the clusters decreased dramatically, leading to an average size of 1.7 nm (Figure 5b and Figure S15b), which is found to be much smaller than the clusters resulting from treatment of the $C_{50}H_{10}$ molecules in air at 300 °C and 400 °C ( Figure S16).

There are several mechanisms that may lead to a reduction in the cluster sizes. For instance, high-temperature-induced sublimation of $C_{50}H_{10}$ molecules out of the clusters, burning and degradation reactions of $C_{50}H_{10}$ molecules when exposed to air and water at high temperatures, and fragmentation and coalescence of the $C_{50}H_{10}$ molecules. The degradation reactions of the $C_{50}H_{10}$ molecules during pretreatment may occur at either the end-cap side, or the open side, or from both sides. These processes can lead to changes not only of the cluster sizes but also of the actual structures of the individual molecules, which is evidenced by nuclear magnetic resonance spectroscopic analysis (NMR, Supporting Information Figure S17). After SWCNT growth, we used AFM to carefully examine tips for many nanotubes. The results show that most SWCNTs do not have larger particles at their tips (Images 1, 2, and 3 in Figure 5c) and that only a small portion of SWCNTs ($< 10\%$) have particles much larger than the diameters of the nanotubes (Image 4 in Figure 5c). This phenomenon is different with a recent study on nanotubes grown from fullerenes where much larger particles were frequently observed at the tips of the nanotubes.[30] We speculate that there is an important difference between $C_{50}H_{10}$ and fullerene, since the latter needs to be opened first to form a cap, which may bring randomization in terms of cluster sizes and structures. AFM characterization shows that the sizes of the seed clusters are pretty small right before nanotube growth (Figure S15b) and that most of the nanotubes possess diameters comparable to those of the seed sizes (Figure 5c). This suggests that most SWCNTs grow from individual (structure changed) molecules or very small aggregates and explains the selective growth of small diameter SWCNTs presented above. Noticeably, in this study we observed the growth of ultra-small SWCNTs, e.g., (5, 4) and (5, 1) (see Figure 2), which are rarely reported under any other nanotube growth process. Typically, a template is needed for the nucleation and growth of such ultra-small SWCNTs, and the grown nanotubes are confined inside the template.[33]

We further used DFT calculations to study how the chirality of nanotubes evolves during their growth process. Conversion of adjacent hexagon-hexagon (6-6) pairs into pentagon-heptagon (5-7) defects is a potential mechanism by which SWCNT chirality would change, as proposed previously by Smalley and Yakobson.[41] Each 6-6 → 5-7 conversion changes a (5, m) SWCNT into a structure that could be a template for the succeeding growth of a (5, m-1) SWCNT. DFT calculations (Figure 6a) indicate a prohibitive barrier impedes this pathway for pristine $C_{50}H_{10}$ molecule. However, the barrier for 6-6 → 5-7 conversion reduces by a factor of three for dehydrogenated $C_{50}H_{10}$ molecules, i.e., from 171.0 kcal/mol (for $C_{50}H_{10}$) to 62.8 kcal/mol (for $C_{50}H_9$, which is the dehydrogenated $C_{50}H_{10}$). Figures 6b and 6c depict the structures of the transition states (TS) for both cases. Since oxygen and water were used during seed pretreatment, it is possible that this could result in dehydrogenation of $C_{50}H_{10}$ or creation of some other active radical species that would have similar low barriers for 6-6 → 5-7 conversion. Such active species would promote a greater extent of chirality change during nanotube growth.

In a similar manner, consecutive 6-6 → 5-7 conversions can take place from the intermediates (or templates) for growth of other (5, m) SWCNTs (m=3, 2, 1) (Figures 6d, 6e, and 6f).[41] DFT calculations further show that these processes are promoted when the 6-6 → 5-7 conversion takes place adjacent to the pentagon of an existing 5-7 defect ( Figures S18, S19, and S20), producing templates for the growth of (5,m-1) SWCNTs. The barriers for these successive transformations are lower than that of the initial (5, 5) →(5, 4) transformation, and reduce slightly in the order: (5, 3) → (5, 2) >(5, 1) →(5, 0) >(5, 2) → (5, 1) > (5, 4) → (5, 3) (Figure 6d). The trend is the same for high temperatures of 1173 K and 298.15 K (Figures S21). The possibility of such transformations to occur can be described as the exponential factor $e^{(-\frac{E_b}{kT})}$, where $E_b$ is the activation free energy (Figure S21), $k$ is the Boltzmann constant, and $T$ is the temperature. At 1173 K (i.e., the nanotube growth temperature used in this study), an energy barrier of $E_b$=60.8 kcal/mol is not very high, suggesting that these steps are not prohibitive for the dehydrogenated species. With more active intermediates, this possibility could be even larger. These theoretical results reveal, in certain degree, the changes of nanotube chirality after growth process. However, the theoretical results cannot explain why there are no (5, 5) SWCNTs grown in the products. The lack of (5, 5) SWCNTs is still very puzzling to us at this stage, which needs further study. In a recent study, Amsharov and Fasel et al. have shown the growth of pure (6, 6) SWCNTs on single crystal Pt substrates by using organic molecules as seeds, under low temperature (400-500 °C) and high vacuum (~$10^{-7}$ mbar) conditions.[28] We note that there are several major differences between their growth approach and ours, including substrates, pressure, and temperature. For example, their CNT growth temperature is much lower than in our method (400-500 °C *versus* 900 °C).

In addition, based on Raman analysis in Figure 3, we observed that some nanotubes have diameters larger than the $C_{50}H_{10}$ molecules, for example, (8, 3). These nanotubes presumably originate from the aggregation of a few molecules into relatively large structures (Figure 5) and consequently, nanotubes with diameters larger than (5, 5) were grown.

Previous theoretical[12, 42, 43] and experimental[44] studies revealed that nanotubes with different chiralities and electronic properties have different stabilities, and metallic nanotubes are generally less stable than semiconducting ones. Researchers have explored such stability and reactivity differences between metallic and semiconducting SWCNTs to realize selective growth of, for example, semiconducting SWCNTs, by introducing external chemical or physical interaction into the CVD environment. Oxidative species like OH radicals,[34] $O_2$ gas,[45] and $H_2O$ vapour[13, 46] were found to be able to preferentially suppress the growth of metallic SWCNTs. Similar phenomena have also been reported by using ultraviolent-assisted CVD.[11] To examine whether a similar mechanism governs our process, we conducted the following experiments.

First, we used a trace oxygen detector to monitor the concentration of oxygen *in situ* during the CVD growth of SWCNTs from pretreated $C_{50}H_{10}$ molecular seeds. The results show that the oxygen concentration ranged from a few ppm to ~77 ppm during the SWCNT growth process (Figure S22). We note that this concentration is two to three orders of magnitude lower than those in the early reports where a few hundred ppm to a few thousand ppm of $H_2O$ and $O_2$ were intentionally added to enable the selective growth of semiconducting SWCNTs.[13, 45, 46] Second, in a separate experiment, we then grew SWCNTs under identical CVD conditions in the absence of $C_{50}H_{10}$ molecular seeds, employing a commonly-used Fe catalyst. Raman analysis shows that SWCNTs grown from Fe have a rather broad diameter distribution and do not show any noticeable enrichment of either metallic or semiconducting SWCNTs (Figure S23). Collectively, the above two experiments reveal that trace oxygen residue in our CVD system does not play an important role for the selective growth of semiconducting SWCNTs and that the selectivity appear to originate from the molecular seeds used.

Previous DFT calculations suggest that structure-dependent stability differences of metallic versus semiconducting SWCNTs are much more significant in the small diameter regime than in medium or large diameter regime.[12] As nanotubes grown from pretreated $C_{50}H_{10}$ end-caps possess exceptionally small diameters, we speculate that nanotube diameter-induced stability differences between metallic and semiconducting SWCNTs, which may play a central role at the very small diameter regime, might be the key reason for the preferential growth of semiconducting SWCNTs in this study. The nucleation and growth of such small diameter SWCNTs on flat substrates may relate to the special molecular seeds used here, which serve as nanotube end-caps and stabilize nuclei of very small diameter nanotubes. Further study is clearly warranted.

In summary, a nanotube-end-cap molecule, $C_{50}H_{10}$, prepared by bottom-up organic chemistry synthesis, was used for the first time to grow SWCNTs having nearly pure semiconducting properties by a metal-free process. Various growth conditions were tested, and their effects on nanotube growth efficiency were studied. The diameter, chirality, and electronic properties of the nanotubes grown from this molecular end-cap were studied in detail via microscopy, spectroscopy, and electrical transport characterization. DFT calculations show that the dehydrogenated $C_{50}H_{10}$ molecules facilitate chirality-changed growth of SWCNTs. The exceptional small diameter feature of the SWCNTs grown from pretreated $C_{50}H_{10}$ molecules, combined with the diameter-dependent stability differences between semiconducting and metallic SWCNTs, are proposed to be the key origin for the nearly exclusive growth of semiconducting SWCNTs. This study not only establishes an efficient approach to grow nearly pure SWCNT semiconductors, but also provides valuable new insight into the selective growth mechanism of SWCNTs.

**Methods**

**Bottom-up synthesis of the $C_{50}H_{10}$ molecular end-cap** In our experiments, $C_{50}H_{10}$ molecules were synthesized from corannulene, which is the smallest curved subunit of $C_{60}$ fullerene and is a bowl-shaped non-planar molecule, with a bowl-depth of 0.87 Å (Figure 1a).[47, 48] It is the oldest known bowl-shaped polycyclic aromatic hydrocarbon (PAH) and was first synthesized in 1966,[49, 50] long before the discovery of fullerenes. Significantly, the synthesis of corannulene has recently been scaled up to produce kilogram quantities,[51] making it by far the most attractive molecular precursor for bottom-up synthesis of nanotubes. Corannulene was subjected to direct chlorination with iodine monochloride, a 5-fold Negishi coupling, and flash vacuum pyrolysis (FVP), successively, to synthesize a hemispherical molecule, $C_{50}H_{10}$,[21] which represents the end-cap plus a short sidewall segment of a (5, 5) chirality SWCNT, as shown in Figure 1b. The as-synthesized $C_{50}H_{10}$ molecules have a nominal purity of 100% without other isomers, and they are quite soluble in common solvents like toluene ($C_6H_5CH_3$), dichloromethane (DCM, $CH_2Cl_2$), and acetonitrile (ACN, $CH_3CN$). The first synthesis of this hemispherical PAH was reported in 2012,[21] and an improved 3-step synthetic route to the same geodesic polyarenes was reported nine months later.[52] More synthesis details can be found in the above two papers.

**Nanotube growth from $C_{50}H_{10}$** The as synthesized $C_{50}H_{10}$ molecules were dissolved in toluene and deposited onto ST-cut quartz substrates via spin-coating (1000 or 2000 rpm for 1 min) or drop-casting, followed by high purity $N_2$ blowing. The quartz substrates were loaded into a 1 inch CVD furnace and subjected to treatment with air at 500 °C for 30 min. Then, a $H_2O$ vapour treatment was conducted at 900 °C for 3 min. During the $H_2O$ treatment, Ar (100 sccm) flowed through a vial containing $H_2O$ kept at 25 °C, and $H_2$ (300 sccm) was also directly introduced into the furnace. After the above pre-treatments, mixed carbon sources ($CH_4/C_2H_4$=1300/10 sccm) together with $H_2$ (300 sccm) were introduced to initiate nanotube growth at 900 °C, typically for 15 min. Lastly, the furnace was cooled down under the protection of 300 sccm $H_2$.

**Nanotube growth from Fe** As control experiments, we also conducted nanotube growth from Fe catalysts. The Fe catalyst stripes were patterned using photolithography and followed by thermal evaporation of 0.3 nm of Fe film. The substrate with Fe film was first annealed at 900 °C for 30 min in air. Then, it was loaded into the growth furnace, and the SWCNT growth followed the identical recipe (temperature, gas flow rates, and gas composition) as with the pretreated $C_{50}H_{10}$ seeds. For CVD growth using different catalysts, we used different quartz reaction tubes and quartz boats, to avoid any cross-contamination between the two catalysts. We also used new quartz tubes and boats in the experiments. Plastic tweezers were used to avoid possible metal residue that may be introduced by metal tweezers. Multiple control experiments using blank quartz (without $C_{50}H_{10}$ molecules) as growth substrate, which underwent the identical pretreatment and growth condition, did not give any CNTs, confirming that there is no contamination in the CVD system.

**Device fabrication and electrical transport measurements** The as-grown SWCNTs on quartz were first transferred onto $Si/SiO_2$ (90 nm) substrates using a PMMA-mediated transfer method.[35] Then, photolithography, e-beam evaporating, and lift-off procedures were conducted to pattern the source and drain electrodes on top of the transferred SWCNTs. The electrodes were Ti/Pd with thicknesses of 1 nm/50 nm, and the channel lengths and widths of the devices vary from 4 μm to 10 μm and 10 μm to 150 μm, respectively. The electrical measurement was conducted on Agilent4156B. The hold time and delay time were set as 2s and 0.2s during measurements.

**Characterization** We used SEM (Hitachi S4700 at an electron accelerating voltage of 1 kV), AFM (Digital Instrument Dimensional 3100, tapping mode), Raman spectroscopy (Renishaw Instrument with laser wavelengths of 633 nm, 514 nm, 457 nm, and 405 nm), and NMR (Varian 400M) to characterize the samples. The oxygen concentration was monitored *in situ* during nanotube growth process using a Trace Oxygen Analyzer (Series 3000, Alpha Omega Instruments). In Raman experiments, the laser spot size was 1-2 μm and the laser power was below 5 mW for all the lasers. The integration time was 30 second or 60 second. We have considered the following three points when assigning a peak to be a RBM. (i) The full width at half maximum intensity (FWHM) of the peak should be >3 cm$^{-1}$,[53] (ii) There should be no less than three data points in the peak, and (iii) The peak should exhibit a good signal-to-noise ratio.

**Calculation details** Local minima and transition state (TS) geometries of $C_{50}H_m$ (m=10, 9, 8, 7, 6, and 5) end-caps with different chirality were optimized at the UB3LYP/6-31G(d) level of theory.[54, 55] All structures were characterized using vibration frequency analyses. All reported electronic energies include zero-point vibration energy (ZPVE) corrections. For the rate calculations, Gibbs free energies were evaluated at temperatures of 298.15K and 1170K under standard pressure. All calculations were performed using the Gaussian 09 package.[56]

*Conflict of Interest.* The authors declare no competing financial interest.

**Acknowledgements** This work was supported by the SRC FCRP FENA center, the Office of Naval Research, the National Science Foundation, and the Air Force Office of Scientific Research. We acknowledge financial support from the National Natural Science Foundation of China (21403127) and the Natural Science Foundation of Shandong Province, China (2014ZRE27498). We also acknowledge Stephen Cronin and Zhe Zhang of University of Southern California for Raman and NMR experiments.

**Author contributions:** B.L, C.Z., and L.T.S. conceived the idea and initiated the study. C.Z. supervised the project. B.L performed nanotube synthesis, SEM, AFM, and Raman characterization, as well as electrical transport measurements. J.L. helped on device fabrication and took part in discussions. E.A.J. and R.B. synthesized the $C_{50}H_{10}$ molecules. H.-B. L., A.P., S.I., and K.M. performed quantum chemical calculations and analysis. B.L. and C.Z. analyzed the electrical transport data. B.L., L.T.S., and C.Z. wrote the manuscript, and all authors commented on the manuscript.

**Supporting Information:**

Additional EDX, Optical absorption, AFM, Raman, device, calculation details, and other results. This material is available free of charge via the Internet at

http://pubs.acs.org.

## Figure legends

**Figure 1 Structure of the molecular end-caps used for nanotube growth.** (a) Structure of the bowl-shaped corannulene molecule ($C_{20}H_{10}$) precursor. (b) Structure of the hemispherical $C_{50}H_{10}$ molecule synthesized from corannulene (a), which represents the end-cap plus a short sidewall segment of a (5, 5) SWCNT. The molecule shown in Figure 1b is used for nanotube growth in this study.

**Figure 2 SEM and AFM characterization of nanotubes grown from $C_{50}H_{10}$ molecular end-caps.** (a) Low magnification SEM image of as grown nanotubes. Inset is a digital camera image of the quartz substrate after deposition of the $C_{50}H_{10}$ molecules and drying, where the red-orange areas correspond to a high density of $C_{50}H_{10}$ molecules. (b), (c), and (e) SEM images of as-grown SWCNTs at the locations indicated in image a. (d) A high magnification SEM image of the area c. (f) An AFM image of a SWCNT with a height of ~0.6 nm.

**Figure 3 Multiple lasers Raman spectroscopic characterization.** (a), (b), (c) Raman RBM spectra of SWCNTs grown from $C_{50}H_{10}$ molecular end-caps excited by 633 nm (a), 514 nm (b), and 457 nm lasers (c). The peaks marked with arrows are from SWCNTs and all the other peaks (marked with *) come from quartz substrates. (d), (e), (f) RBM frequency distributions based on the above three lasers. (g) Diameter distribution of SWCNTs derived from the RBM frequencies using the equation $d_t = \frac{223.5}{\omega_{RBM} - 12.5}$. (h) Raman D-band and G-band spectra of SWCNTs excited by a 457 nm laser.

**Figure 4 Electrical transport property and breakdown of SWCNT FETs.** (a) Schematic of device structure of a back-gated individual SWCNT FET. (b) Statistics of on/off current ratio distribution of 34 individual SWCNT FETs. (c), (d) Representative transfer characteristics ($I_{DS}$-$V_G$) of an individual semiconducting (c) and semi-metallic (d) SWCNT FET with the SEM images of the devices shown in the inset. (e) Transfer characteristic of an all-semiconducting nanotube-array-FET, with the inset SEM image showing a total of four SWCNTs connected to both electrodes. (f) Electrical breakdown experiments of the device in e. (g) Transfer characteristics of a multiple-nanotube-FET before (blue) and after (red) electrical breakdown, with the inset SEM image showing a total of two SWCNTs connected to both electrodes. (h) Electrical breakdown experiments of the device in g. Scale bar of the SEM images: 5 μm for b and d, 10 μm for e, and 20 μm for g.

**Figure 5 Molecular seed size evolution and nanotube diameter-seed size relationship.** (a), (b) AFM images of as-deposited $C_{50}H_{10}$ molecular aggregates on quartz (a) and the $C_{50}H_{10}$ molecules after pretreatment (b). (c) AFM examinations of the relationship between as-grown nanotubes and the seed molecules. Images 1, 2, and 3 show an end of one nanotube, two ends of two different nanotubes, and both ends of one nanotube, respectively. No big particles were found at the nanotube ends. Image 4 shows a big particle at the end of one SWCNT. The ends of the SWCNTs are indicated by green arrows. The vertical bars are 15 nm for all AFM images.

**Figure 6 DFT calculations of the energy profiles of the (5,m) → (5,m-1) chirality transformations and the structures of the transition states (TS).** (a) Energy barriers of (5, 5)→(5, 4) transformation for $C_{50}H_{10}$ and $C_{50}H_9$. (b), (c) Structures of TS of the (5, 5) →(5, 4) transformation starting from $C_{50}H_{10}$ and $C_{50}H_9$, respectively. (d) Energy barriers of (5, m)→(5, m-1) transformations for m=5, 4, 3, 2, and 1, starting from $C_{50}H_{m+4}$. (e),(f) Structures of the formed (5, 4) and (5, 3) chirality, with one and two adjacent 5-7 pairs, respectively. The calculated energy includes the zero point vibration energy (ZPVE).

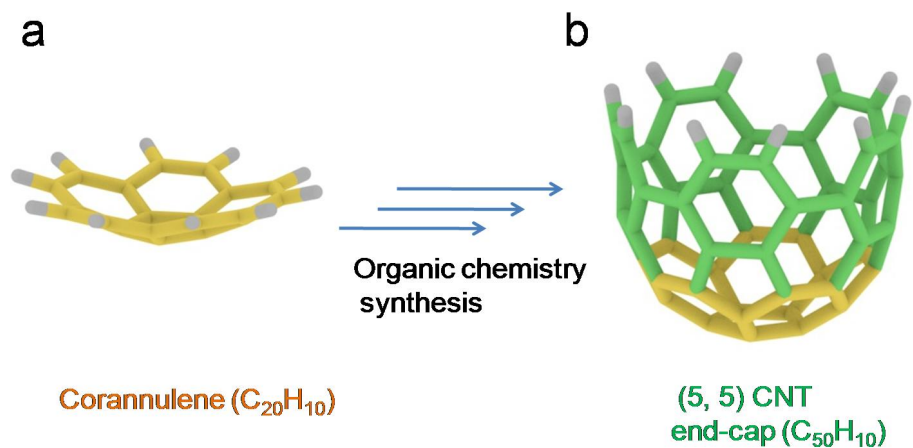

a

b

Organic chemistry
synthesis

Corannulene (C$_{20}$H$_{10}$)

(5, 5) CNT
end-cap (C$_{50}$H$_{10}$)

Figure 1

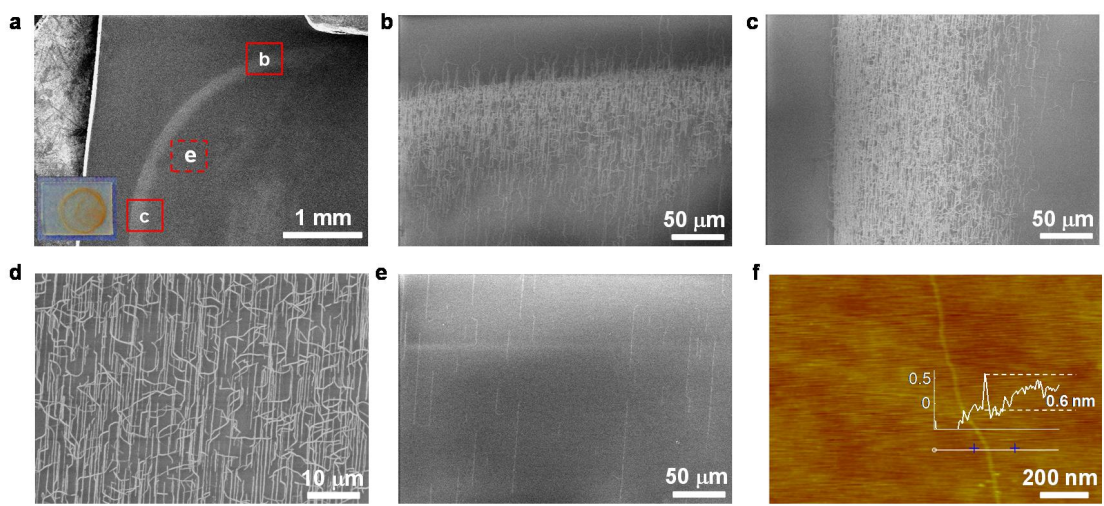

Figure 2

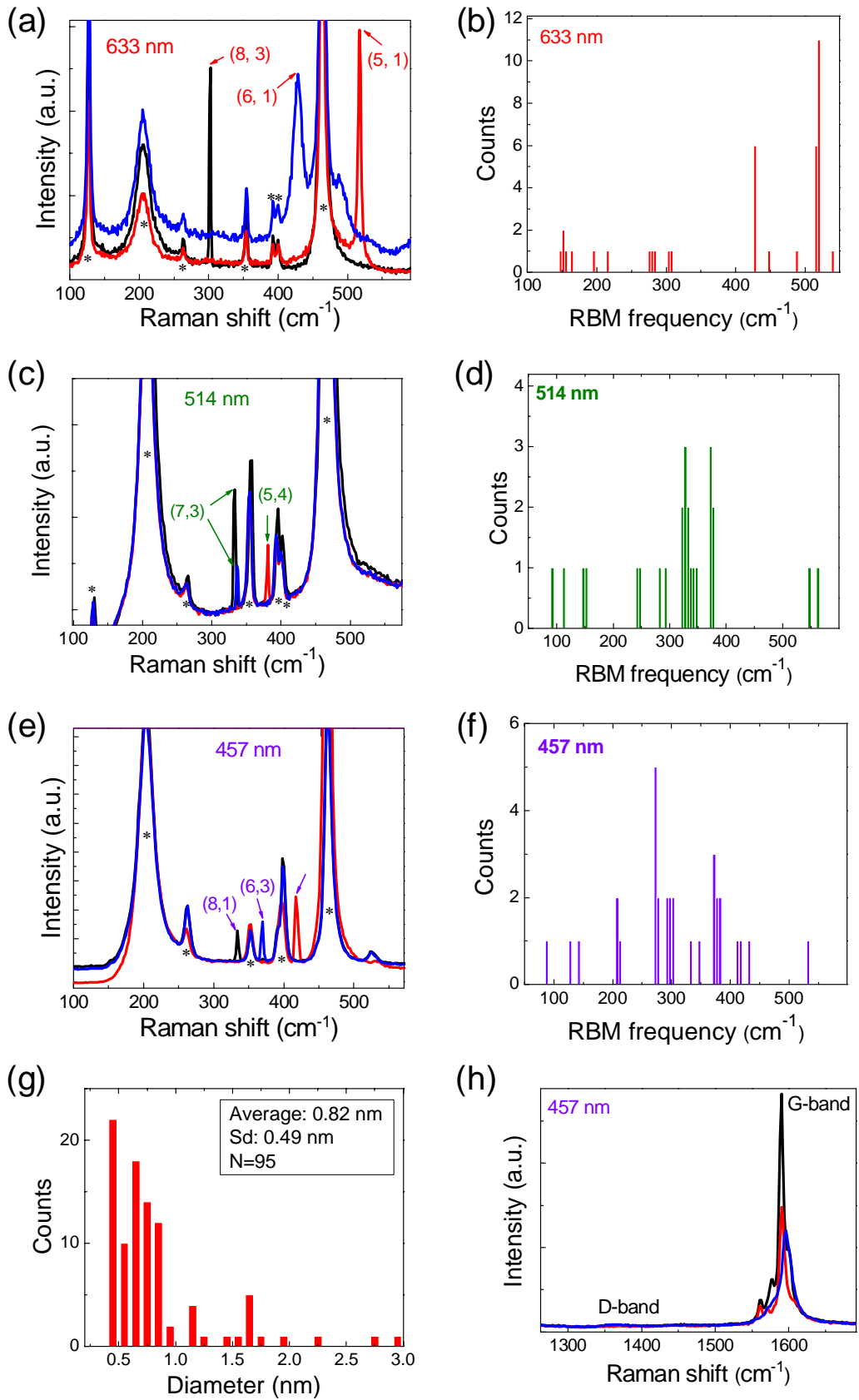

Figure 3

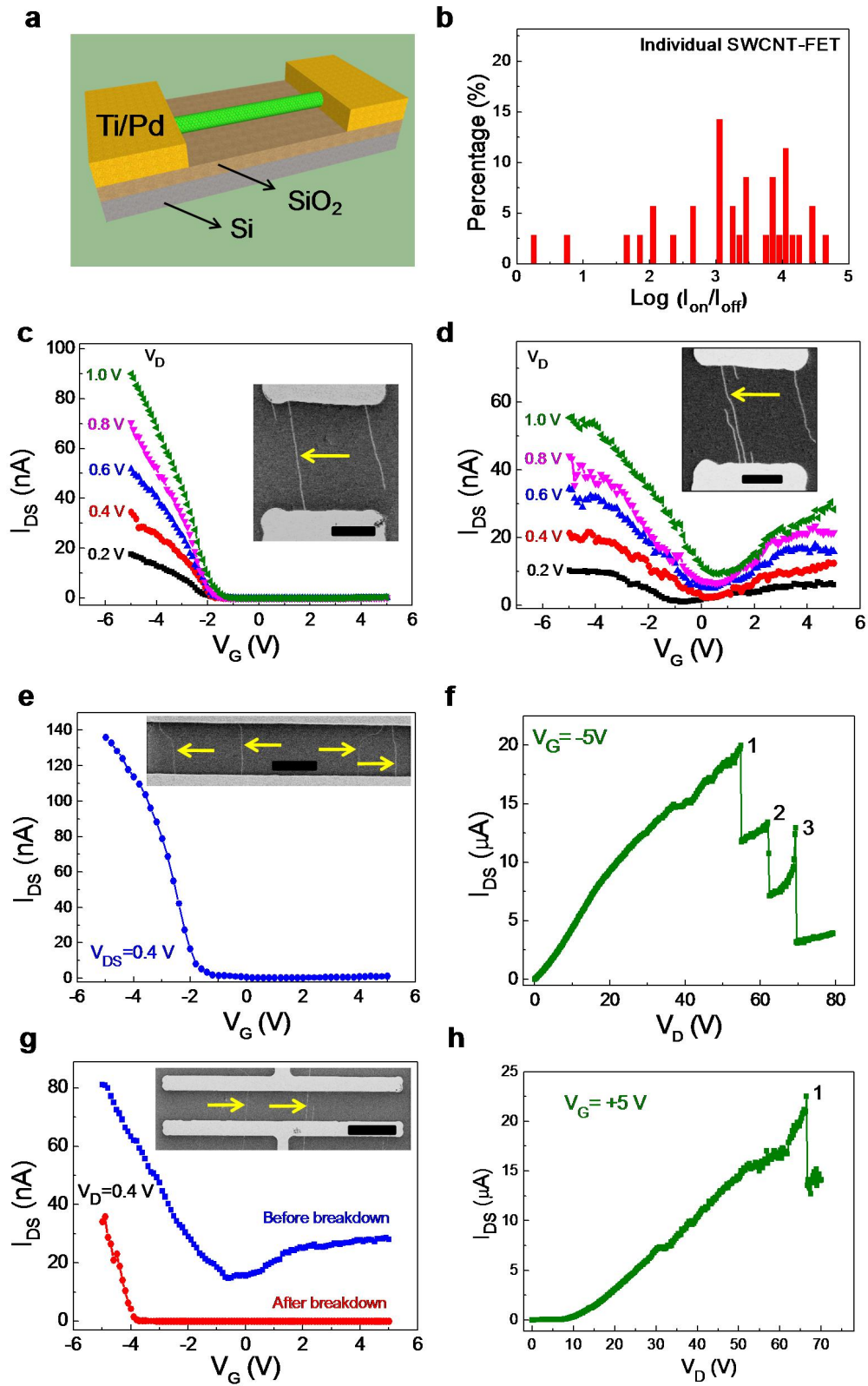

Figure 4

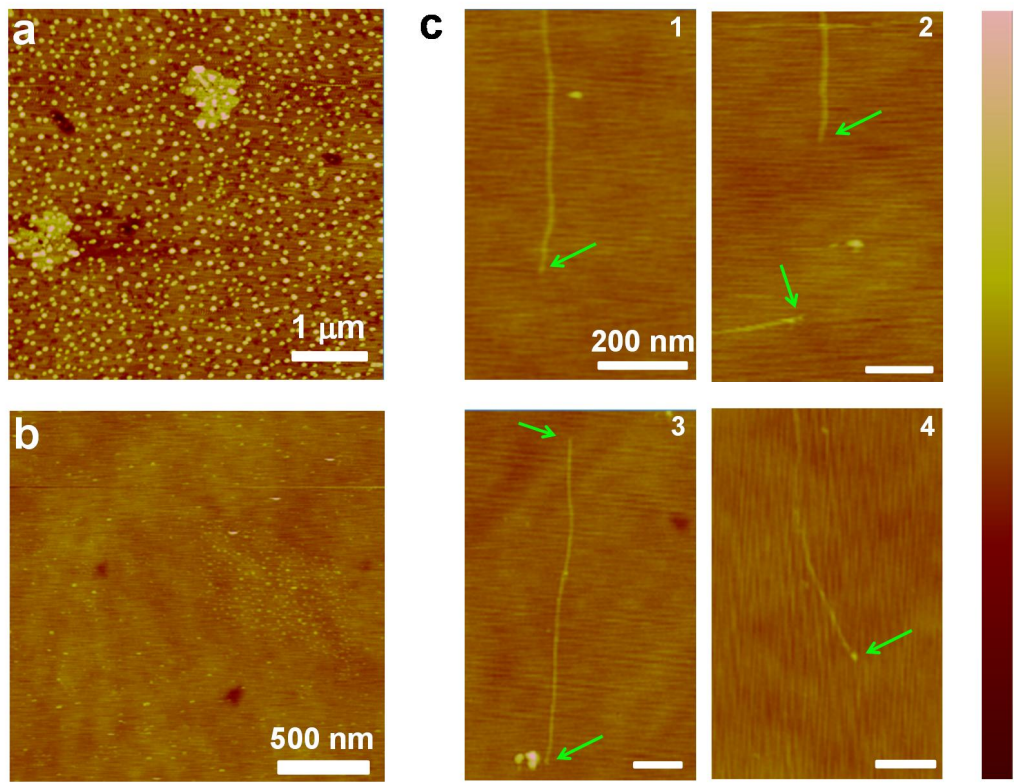

Figure 5

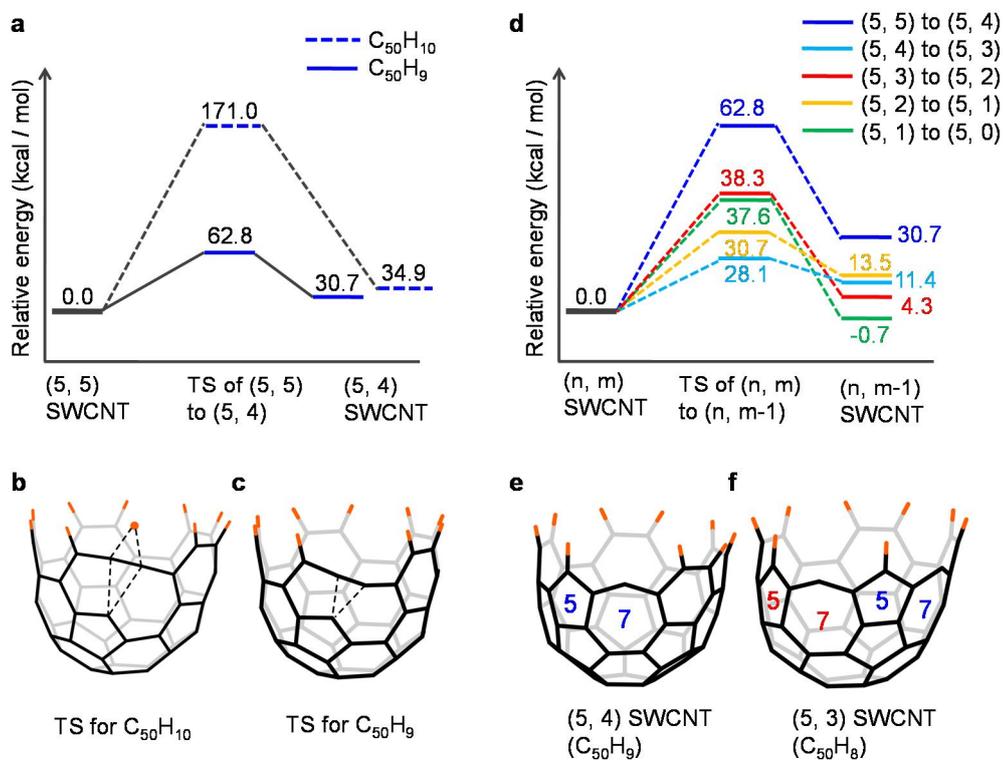

Figure 6

**TOC image.**

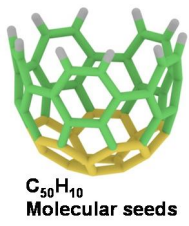

$C_{50}H_{10}$
Molecular seeds

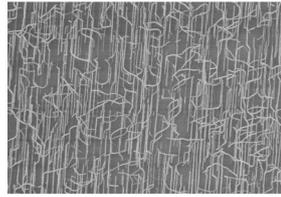

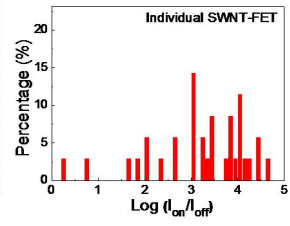